\documentclass[twoside,superscriptaddress,twocolumn,floatfix,a4paper,prl,longbibliography]{revtex4-1}
\usepackage{color}
\usepackage{graphicx}
\usepackage[utf8x]{inputenc}
\usepackage[T1]{fontenc}
\usepackage{siunitx}
\usepackage{amstext}
\usepackage{hyperref}
\usepackage{epstopdf}
\usepackage{amsfonts}
\usepackage{amsmath}
\usepackage{multirow}
\usepackage{tabularx}
\usepackage{filecontents}

\newcommand{\mdf}[1]{{#1}}

\begin{document}

\title{Experimental kernel-based quantum machine learning in finite feature space}

\author{Karol Bartkiewicz} \email{karol.bartkiewicz@upol.cz}
\affiliation{Faculty of Physics, Adam Mickiewicz University, 
PL-61-614 Pozna\'n, Poland}
\affiliation{RCPTM, Joint Laboratory of Optics of Palacký University 
and Institute of Physics of Czech Academy of Sciences, 17. listopadu 12, 
771 46 Olomouc, Czech Republic}

\author{Clemens Gneiting}
\affiliation{Theoretical Quantum Physics Laboratory, RIKEN Cluster for 
Pioneering Research, 351-0198 Wako-shi, Japan}

\author{Anton\'{i}n \v{C}ernoch}
\affiliation{RCPTM, Joint Laboratory of Optics of Palacký University and 
Institute of Physics of Czech Academy of Sciences, 17. listopadu 12, 
771 46 Olomouc, Czech Republic} 

\author{Kate\v{r}ina Jir\'{a}kov\'{a}}
\affiliation{RCPTM, Joint Laboratory of Optics of Palacký University and 
Institute of Physics of Czech Academy of Sciences, 17. listopadu 12, 
771 46 Olomouc, Czech Republic}

\author{Karel Lemr}
\email{k.lemr@upol.cz}
\affiliation{RCPTM, Joint Laboratory of Optics of Palacký University and 
Institute of Physics of Czech Academy of Sciences, 17. listopadu 12, 
771 46 Olomouc, Czech Republic} 

\author{Franco Nori}
\affiliation{Theoretical Quantum Physics Laboratory, RIKEN Cluster for 
Pioneering Research, 351-0198 Wako-shi, Japan}
\affiliation{Department of Physics, The University of Michigan, Ann Arbor, MI 48109-1040, USA}

\begin{abstract}
We implement an all-optical setup demonstrating kernel-based quantum machine learning for two-dimensional classification problems. In this hybrid approach, kernel evaluations are outsourced to projective measurements on suitably designed quantum states encoding the training data, while the model training is processed on a classical computer. Our two-photon proposal encodes data points in a discrete, eight-dimensional feature Hilbert space. In order to maximize the application range of the deployable kernels, we optimize feature maps towards the resulting kernels' ability to separate points, i.e., their ``{\it resolution},'' under the constraint of finite, fixed Hilbert space dimension. Implementing these kernels, our setup delivers viable decision boundaries for standard nonlinear supervised classification tasks in feature space. We demonstrate such kernel-based quantum machine learning using specialized multiphoton quantum optical circuits. The deployed kernel exhibits exponentially better scaling in the required number of qubits than a direct generalization of kernels described in the literature. 
\end{abstract}

\date{\today}

\maketitle

\paragraph*{Introduction.}
Many contemporary computational problems like drug design, traffic control, 
logistics, automatic driving, stock market analysis, automatic medical 
examination, material engineering, and others, routinely require optimization over huge 
amounts of data~\cite{MLbook}. These highly demanding problems can be approached 
by suitable machine learning (ML) algorithms. However, in many relevant cases the underlying calculations last prohibitively 
long. These computations could potentially run more efficiently
(sometimes exponentially faster) by utilizing quantum resources in ML
algorithms (i.e., QML). This speed-up can be partially attributed to 
the collective processing of quantum information mediated by quantum 
entanglement. There are various approaches to QML that could be characterized as
linear algebra solvers, sampling, quantum optimization, or using quantum circuits 
as trainable models for inference (see, e.g., Refs. \cite{schuld2015introduction, Cai2015PRL,
Biamonte2017Nature, Schuld2017EPL,Gao2018PRL, ciliberto2018quantum,Rebentrost2014quantum,PhysRevLett.114.140504,Chatterjee2017,Schuld2019PRL,havlivcek2019supervised}). Most 
of the focus both in classical ML and in QML has been put on deep learning
and neural networks. However, recently a promising {\it kernel-based} approach to 
supervised QML has been proposed in \cite{Chatterjee2017,Schuld2019PRL}. 
It is especially interesting consider its implementation on the platform of linear optics as it does not require quantum memory, but rather  combining classical and quantum 
computations. We theoretically elaborate this kernel-based QML (KQML) for multiphoton quantum optical circuits using a kernel that exhibits 
exponentially better scaling in the number of required  qubits than a 
direct generalization of kernels previously discussed in the literature. We implement
this scheme in a proof-of-principle experiment.

\begin{figure}
	\includegraphics[width=8.0cm]{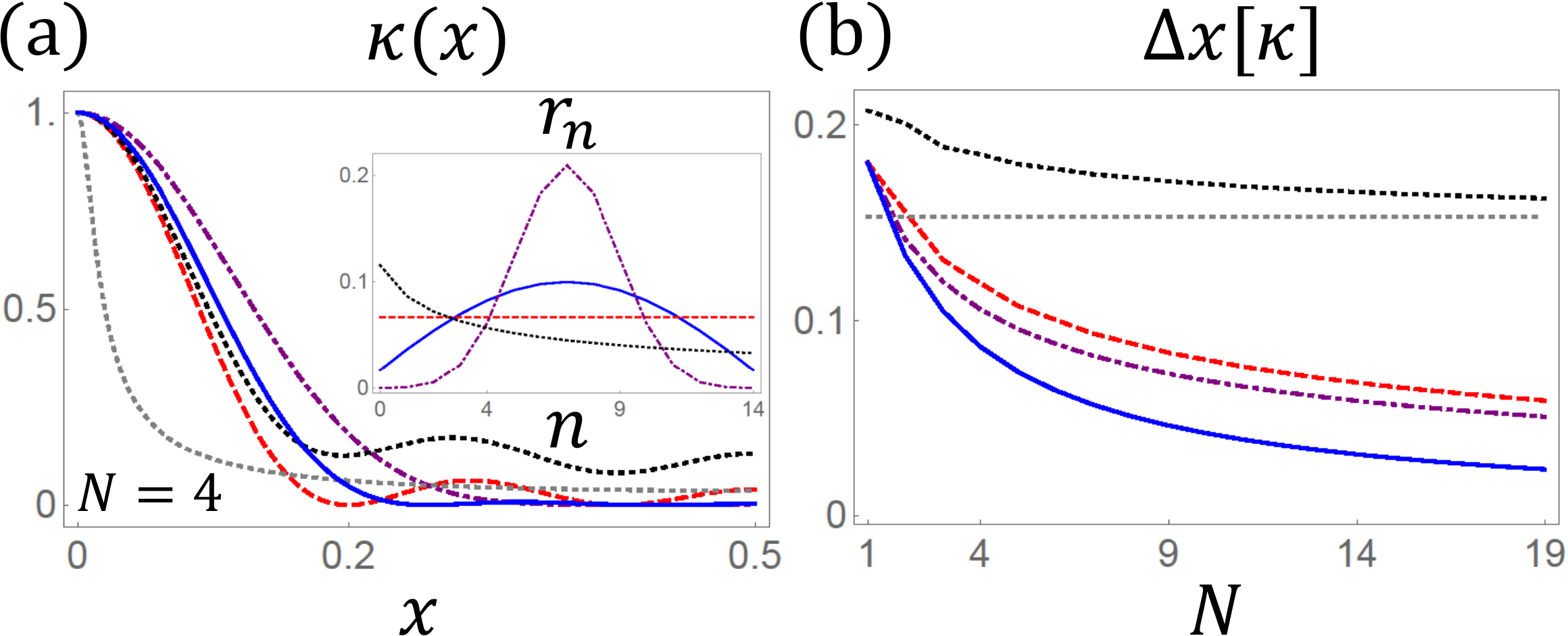}
	\caption{\label{fig:1} \mdf{Kernel family (\ref{Eq:modular_squeezing_kernel}) for different amplitude choices. (a) We find that the resolution-optimized kernel (blue solid) exhibits suppressed side maxima as compared to the MSI kernel (red dashed), while the TSQ kernel (with squeezing factor $
	\zeta=2$, black dotted) maintains a nonvanishing plateau at all $x$ values. For comparison, we also display the respective squeezed-state kernel for $N \rightarrow \infty$ (gray dotted) and CK (purple dash-dotted). The inlay shows the characteristic amplitude progressions for the example $N=14$ and $\zeta=4$. (b) The optimized kernel exhibits a significantly improved resolution progression with $N$ as compared to the MSI or the TSQ kernel (here with $
	\zeta=3$).}}
\end{figure}

\begin{figure}
\includegraphics[width=8.5cm]{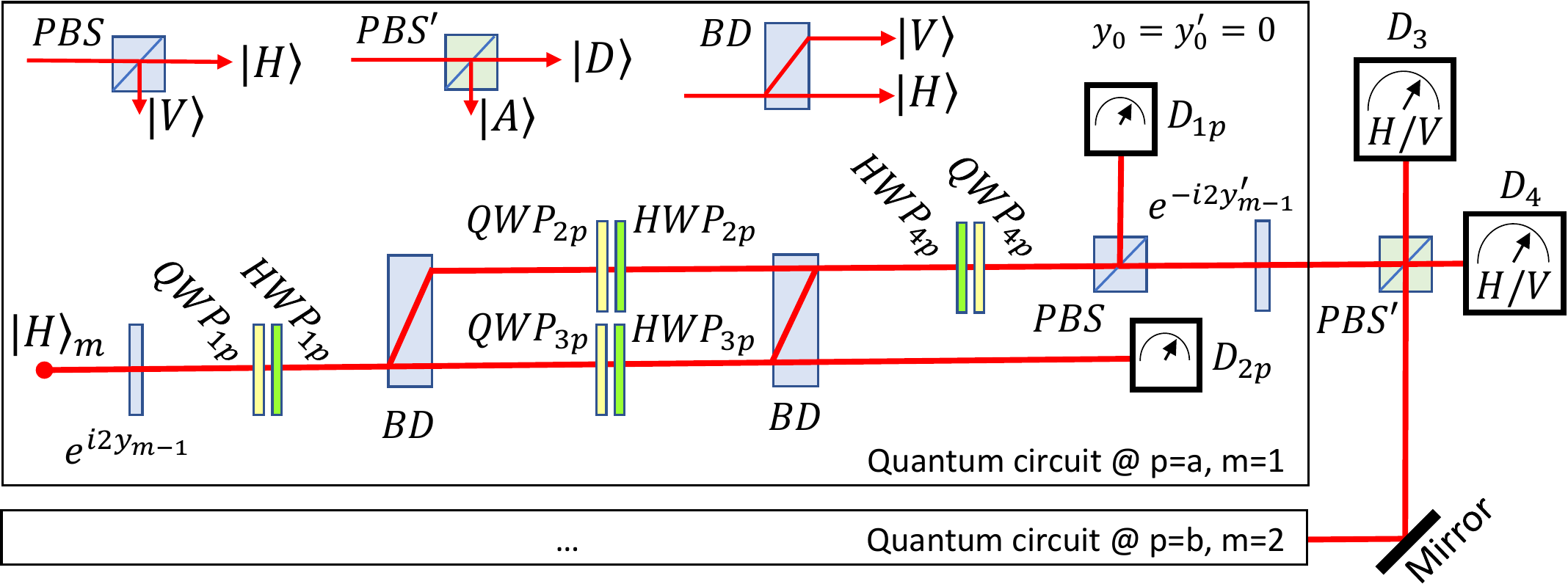}
\caption{\label{fig:2} Optical circuit implementing both the FM 
and the model circuits. The performance of the setup in QML
is shown in Fig.~\ref{fig:3} for $N=1$ and $D=2.$  The experimental 
setup consists of polarizing beam splitters ($PBS$s), 
beam dividers ($BD$), quarter-wave and half-wave plates ($QWP$s and $HWP$s, 
respectively), and single photon detectors $D_{n}$ for $n=1a,1b,2a,2b,3,4$.
$D_3$ and $D_4$ are $H/V$ polarization 
resolving (implemented as a $PBS$ and two standard detectors). 
The kernel $\kappa(x',x)_{\mathrm{exp}} =[\sum_{p,s=H,V} 
CC(D_{2s},D_{3p}) - CC(D_{2V},D_{3H})+CC(D_{2H},D_{3V})]/
\sum_{m>n}\sum_{n=1}^6 CC(D_{m},D_{n})$
is given as a ratio of coincidences $CC(D_{m},D_{n})$ 
registered by photon detectors $D_{n}$ and $D_{m}$ to 
the total number of photons.}
\end{figure}

Let us explain KQML by first recalling some 
definitions and theorems and then we overview the recently proposed method for 
finding linear boundaries in feature Hilbert space (FHS)~\cite{Schuld2019PRL}.
FHS is defined as a space of complex vectors 
$|\varphi(x)\rangle,$ where $\varphi$ is a feature map (FM) and $x$ is a real vector of 
dimension $D$ (the input data). FHSs generally have higher 
dimension than the original data $x$. This implies that the linear decision 
boundary in FHS can give rise to a nonlinear decision boundary 
in the original data space. By virtue of such nonlinear FMs, we do not 
need to implement nonlinear transformations on the directly encoded data $x,$ which 
is usually encoded as amplitudes in other QML approaches. Another benefit of 
KQML is that we can directly measure inner products of vectors mapped onto 
FHS. Thus, we are able to physically measure a kernel function 
$\kappa(x',x)=|\langle \varphi(x')|\varphi(x)\rangle|^2,$
instead of computing it. This could, in some cases, be much faster than the latter
option. It follows from the representer theorem that any function of the 
reproducing kernel that minimizes the cost function (the solution to the ML problem) 
can be written as 
$f^*(x)=\sum_{m=1}^M a_m \kappa(x,x^m),$ where $M$ is the number of training samples, and $a_m$ are some real parameters
subject to the training, and $x$ belongs to feature space. Once the kernel $\kappa$ is known, the parameters 
$a_m$ can be found very efficiently. 
The goal of the ML is to find a function $f^*(x)$ that classifies the non-separable 
points ${x^1,...,x^{M-K}}$ and ${x^{M-K+1},...,x^M}$ by finding a trade-off
between the number of misclassifications and the width of the separating
margin. The parameters $a_m$ can be obtained by solving the following problem:
minimize    $\sum_{m=1}^M (|a_m|^2 + \gamma u_m)$
s.t.    $a_i \kappa(x,x^i) \geq 1 - u_i$        for $i = 1,...,M-K$ and
        $a_i \kappa(x,x^i) \leq -(1 - u_i)$     for $i = M-K+1,...,M,$
$u \geq 0,$ where $\gamma$ gives the relative weight of the number 
of misclassified points compared to the width of the margin.
In a nutshell, this approach allows to both 
replace the nonlinearity of the problem with linear multidimensional quantum computations, which offers a potential speed-up.

\paragraph{Kernel resolution in finite dimensions.}
The ability of a kernel to distinguish data points (i.e., its {\it resolution}) is an essential hyperparameter, which, for given training data, can decide if a model can be trained successfully or not. 
If kernel resolution is too coarse, resulting decision boundaries miss relevant details in the data, if it is too refined, the model becomes prone to overfitting. In the infinite-dimensional feature spaces offered by continuous variable implementations, viable FMs with adjustable resolution can be implemented, e.g., by mapping data into squeezed states \cite{Schuld2019PRL}, where the  adjustable squeezing factor determines the resolution. 

Within the paradigm of discrete, finite-dimensional quantum information processing, FHS dimension becomes a scarce resource, resulting in limitations on kernel resolution. Let us discuss the optimal kernel resolution that can be achieved in $N$-dimensional FHS, within the class of FMs of the form
\begin{align} \label{Eq:Interference_states}
x \rightarrow |\psi(x)\rangle = \sum_{n=0}^{N} \sqrt{r_n} e^{2 \pi i n x} \, |n\rangle \hspace{3mm} , \hspace{3mm} \sum_{n=0}^{N} r_n = 1 ,
\end{align}
with $\{|n \rangle \}$ a basis of the Hilbert space and $x \in [-1/2, 1/2)$. Any data set can be brought to this form, which is a routine step in data preparation. We stress that the amplitudes $r_n$ are independent from the input values $x$. The resulting kernels then are of the form
\begin{align} \label{Eq:modular_squeezing_kernel}
\kappa(x,x') = \kappa(x-x') = \big{|}\sum_{n=0}^{N} r_n \, e^{2 \pi i n (x'-x)}\big{|}^2 .
\end{align}
In this shorthand notation $\kappa(x) \geq 0 \, \forall x$ and $\kappa(0)=1$. For the sake of clarity we consider here 1D input data $x$. For $D$-dimensional inputs $\vec{x}$, each input component $x_i$ is encoded separately, requiring an $(N\cdot D+D)$-dimensional FHS. If the FHS is spanned by $q$ qubits, we have $N=2^q-1$.  In particular, for $N=1$ and $r_n = 1/2$ we have $\kappa(x,x')=\cos[\pi(x'-x)]^2,$ which realizes a {\it cosine kernel} (CK). The class of states (\ref{Eq:Interference_states}) comprises also truncated squeezed states $|\psi_{\rm TSQ}(x)\rangle$, with $\sqrt{r_n} = \frac{\sqrt{(2 n)!} (-\tanh \zeta)^n}{\sqrt{B} \, 2^n n! \sqrt{\cosh 
\zeta}}$ ($
\zeta$ denotes the squeezing factor and $B$ renormalizes the state after truncation), and, what we call here, {\it multi-slit interference states} $|\psi_{\rm MSI}(x)\rangle$, with constant amplitudes $\sqrt{r_n} = 1/\sqrt{N}$. The latter inherit their name from the fact that, by virtue of $\langle x|p \rangle = e^{2 \pi i p x}$ ($h$=1), they can be related to a balanced superposition of momentum states in a compact continuous variable Hilbert space (augmented by an internal spin-$N$ degree of freedom), $|\psi_{\rm MSI}(x)\rangle = \frac{1}{\sqrt{N}} \sum_{n=1}^{N} \langle x| p=n \rangle |n\rangle$, giving rise to ``$N$-slit interference'' in the position coordinate when projected onto $\langle x| \otimes \frac{1}{\sqrt{N}} \sum_{n=1}^{N} \langle n|$ \cite{Gneiting2011detecting}. Note that polynomial kernels (discussed, e.g., in \cite{Rebentrost2014quantum, Schuld2019PRL}) fall outside of the state class (\ref{Eq:Interference_states}).

We can use the above compact-space embedding to gain further insight into the nature of our kernel definition (\ref{Eq:modular_squeezing_kernel}). If we interpret the states (\ref{Eq:Interference_states}) as $|\psi\rangle = \sum_{n=1}^{N} \sqrt{r_n} |p=n\rangle \otimes |n\rangle$, we can introduce the density operator $\rho = |\psi\rangle \langle \psi|$ and trace over the internal spin degree of freedom, $\rho_{\rm ext} = {\rm Tr}_{\rm int} \rho = \sum_{n=1}^{N} r_n |p=n\rangle \langle p=n|$. We then find that the kernel (\ref{Eq:modular_squeezing_kernel}) is related to the spatial coherences of the mixed reduced state $\rho_{\rm ext}$: $\kappa(x,x') = |\langle x|\rho_{\rm ext}|x' \rangle|^2$.

We define a kernel's spatial resolution $\Delta x[\kappa]$ by its spatial variance
\begin{align} \label{Eq:kernel_definition}
(\Delta x[\kappa])^2 \equiv \int_{-1/2}^{1/2} dx \, x^2 \tilde{\kappa}(x) ,
\end{align}
where the renormalized kernel $\tilde{\kappa}(x) = \kappa(x)/R$, with $R \equiv \int_{-1/2}^{1/2} dx \, \kappa(x) = \sum_{n=1}^{N} r_n^2$, describes a valid probability distribution. In the case of the mulit-slit interference states $|\psi_{\rm MSI}\rangle$, one analytically obtains $(\Delta x[\kappa_{\rm MSI}])^2 = \frac{1}{12} (1-S_1(N))$, with the interferometric ``squeezing factor'' $S_1(N) = -\frac{12}{\pi^2} \sum_{j=1}^{N-1} (-1)^j \frac{N-j}{N j^2}$, and $N \geq 2$ \cite{Gneiting2011detecting}.

The kernel (\ref{Eq:kernel_definition}) minimizing the resolution is a solution to the optimization problem: minimize $\frac{\vec{r}^T \cdot K \cdot \vec{r}}{|\vec{r}|^2}$ s.t. $\sum_{n=1}^{N} r_n = 1$, where
\begin{align}
K_{n m} = \left\{
\begin{array}{ll}
\frac{1}{12}, & n=m \\
\frac{(-1)^{|n-m|}}{2 (n-m)^2 \pi^2}, & \, \textrm{else} \\
\end{array}
\right.
\end{align}
and $\vec{r}=(r_1, \dots, r_N)^T$. In Figure~\ref{fig:1} we compare this optimized kernel with the TSQ and the MSI kernel. The optimized kernel comes with strongly suppressed side maxima as compared to the MSI kernel, while the TSQ maintains a nonvanishing plateau for all $x$ values. Consequently, the optimized kernel enables, for a given $N$, a significantly improved resolution as compared to the other kernel choices. The inlay of Figure~\ref{fig:1}a clarifies that amplitudes decaying symmetrically about the ``center'' state lie at the heart of the performance advantage in terms of kernel resolution.

Although kernels of the form (\ref{Eq:modular_squeezing_kernel}) can also be efficiently computed classically, their quantum evaluation may still deliver a significant speed-up. Moreover, seen as a module to be combined with other subroutines, the FMs proposed here may contribute its resource-efficient data point separation ability to an overall setup that comes with an inherently quantum scaling advantage. MSI states, for instance, can be generated in a gate-based quantum computer following the first stage of the phase-estimation algorithm \cite{nielsen2002quantum}.

\paragraph*{Cosine kernels.}
The kernel for our proof of principle demonstration of KQML is defined as 
$\kappa(x',x)=|\langle \varphi(x')|\varphi(x)\rangle|^2=\prod_{n=1}^{D}
\cos^{2N}(x_n'-x_n),$ where the FM taking a normalized feature $x_n\in[-\pi/2,\pi/2)$ to FHS
is $|\varphi(x)\rangle =\bigotimes_{n=1}^{D} \sum_{k=0}^{N} \sqrt{N\choose k}
\sin^k(x_n)\cos^{N-k}(x_n)|k\rangle_n.$ 
Note that $N$ is related to the number of qubits $q$ per dimension 
as $q=\lceil \log_2(N+1)\rceil .$ 
This FM can also be considered a constant-phase 
representation of constant-amplitude states. This is the same as representing states either in a basis of eigenstates of $x$ or $z$ components of a collective spin operator. In particular,   
$(\cos(x)|0\rangle + \sin(x)|1\rangle)/\sqrt{2}\Leftrightarrow(|0'\rangle + e^{2ix}|1'\rangle)/\sqrt{2},$
where $|0\rangle = (|0'\rangle+|1'\rangle)/\sqrt{2}$ and $|1\rangle = (|0'\rangle-|1'\rangle)/\sqrt{2}.$

This mapping uses  less 
resources than the direct product of the map from Ref.~\cite{Schuld2019PRL}, i.e., 
$|\varphi(x)\rangle =\bigotimes_{n=1}^{D} \bigotimes_{m=1}^{N}
\sum_{k=0}^1\sin^k(x_n)\cos^{1-k}(x_n)|k\rangle_{n,m},$ 
where the number of qubits per dimension is $q=N.$ Using 
the powers of CKs allows us to adjust the kernel resolution by 
choosing the proper value of $N.$ Thus, the number of used qubits can be related 
directly to the spread of the kernel. The number of 
qubits here plays the same role as the squeezing parameter in the experimental 
proposal given in Ref.~\cite{Schuld2019PRL}.
The CK can also include additional $(D-1)$ degrees of freedom by virtue of a FM defined as
\begin{equation}\label{eq:fmapb}
|\varphi(x)\rangle =\bigotimes_{n=1}^{D} \sum_{k=0}^{N}e^{i2y_{n-1}} 
\sqrt{N\choose k}\sin^k(x_n)\cos^{N-k}(x_n)|k\rangle_n,
\end{equation}
where $y_0=0,$ the number of terms here is $(N+1)^D,$
and the associated kernel measured by postelection  is
$\kappa(x',x)=\prod_{n=1}^D\cos^{2N}(x'_n-x_n)\cos^2(y'_{n-1}-y_{n-1}).$

\paragraph*{Optical circuit for KQML.}
States given by Eq.~(\ref{eq:fmapb}) can be prepared in a quantum optical setup. 
In the reported proof of principle experiment, we can set 
$N=3$ and $D=2$. This means that, effectively, the experiment deploys $q=2$ 
qubits per dimension. The FM is defined via single-photon 
polarization states ($H/V$ polarization) as well as dual-rail encoding 
($T/B$ for top/bottom rail, respectively)
\begin{eqnarray}
|\varphi(x)\rangle &=& \bigotimes_{n=1}^2 \left(c^3(x_n)|HT\rangle_n 
+ \sqrt{3}s(x_n) c^2(x_n)|HB\rangle_n\right. \nonumber \\
&&\left. + \sqrt{3}c(x_n) s^2 (x_n)|VB\rangle_n + s^3(x_n)|VT\rangle_n \right), 
\end{eqnarray}
where $c(x_n)\equiv\cos(x_n)$ and $s(x_n)\equiv\sin(x_n).$ This approach is 
resource-efficient as it only requires two photons to encode $x$ into the FHS
state of $N=3$ and $D=2$. An optical circuit implementing this FM 
is depicted in Fig.~\ref{fig:2}.  The top part of the FM circuit 
works as follows: first, it transforms the standard input $|HB\rangle$ using wave 
plates resulting in $|HB\rangle\to (|HB\rangle + |VB\rangle)/\sqrt{2}.$ Next, 
a beam divider separates polarization modes in space, i.e., we have $(|HB\rangle 
+ |VT\rangle).$ Now, the effective operation of wave plates in the top and bottom 
modes can be described as first transforming $|VT\rangle \to \mu_T|HT\rangle 
+  \nu_T|VT\rangle$ and $|HB\rangle \to \mu_B|HB\rangle +  \nu_B|VB\rangle.$ 
The parameters are set as 
$\mu_T = \sqrt{2}c^3(x_n),$ $\nu_T = \sqrt{2}{s^3(x_n)},$ $\mu_B = 
\sqrt{6}c^2(x_n)s(x_n),$ $\nu_B = \sqrt{6}c(x_n)s^2(x_n).$ 
This whole operation is unitary and can be described as $U(x)|HH\rangle = 
|\varphi(x)\rangle.$ The complex conjugate of operation $U(x)$ is  
$U^\dagger(x')$ and it can be used to express the kernel as
$\kappa(x',x) = |\langle HH|U^\dagger(x')U(x)|HH\rangle|^2.$
Thus, the circuit  $U^\dagger(x')$ 
for projecting the state $|\varphi(x)\rangle$ to $|\varphi(x')\rangle$ can be 
constructed as the inverse of the feature embedding  $U(x)$
circuit, but for setup parameters set for $x'$. 
The next action of the plates in the top and bottom 
rails is to perform a reverse transformation , but for 
$x_n=x'_n.$ Next,  the plates  flip the 
polarizations in the respective rails. Now, the interesting part of the 
engineered state is in the top rail with flipped polarization. To implement 
$U(x')^\dagger,$ the last pair of waveplates is used both to flip the 
polarization and to perform the Hadamard transformation. 
Finally, the $PBS$ transmits only $H$-polarized 
photons for further processing.
The procedure of measuring the kernel $\kappa(x',x)$ can be extended to include  
additional dimensions, resulting in measuring the 
kernel $\bar\kappa(x',x)=\kappa(x',x)\cos^2(y-y')$
following from FM (\ref{eq:fmapb}). Instead of the transformation 
$U^\dagger(x')U(x),$ we consider
$R^\dagger(y')U^\dagger(x')U(x)R(y),$
where $R(y)= e^{2iy}|H\rangle\langle H|$ is a phase shift applied to 
a preselected $H$-polarized photon in the bottom part of the setup, and  
$R^\dagger(y')= e^{-2iy'}|H\rangle\langle H|$ is a phase shift to the
postselected $H$-polarized photon in the same part of the setup. 
The phase difference between the postselected upper and lower 
$H$-polarized photons can be measured as $\cos^2(y-y').$ 
This is done with $PBS'$ which transmits 
diagonally-polarized photons $|D\rangle = (|H\rangle+|V\rangle)/\sqrt{2}$ 
and reflects antidiagonal photons $|A\rangle = (|H\rangle-|V\rangle)/\sqrt{2},$  
and polarization-resolving single-photon detectors (see caption
of Fig.~\ref{fig:2}).

\begin{figure}
\includegraphics[width=8.0cm]{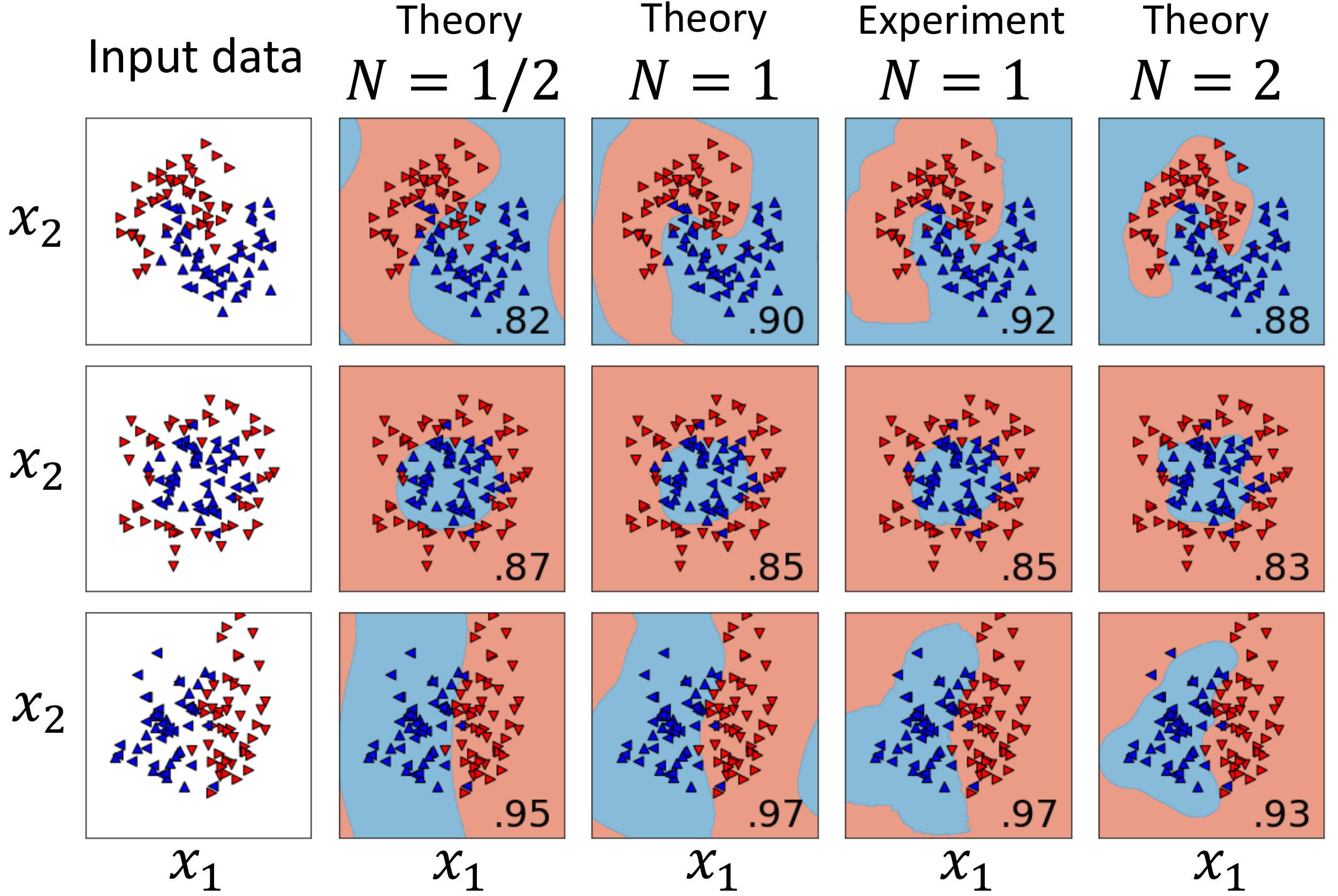}
\caption{\label{fig:3} Training results on a random inseparable data set of 
40 samples (up/down-tipped triangles). The performance on a test set 
(left/right-tipped triangles)
of 60 points (the fraction of 
correctly classified samples that were not used in the QML process) is given 
in the bottom right corner of each respective subplot.  
We see that the best choice of CK is $N=1$. For $N=2$ we deal with 
overfitting and for $N=1/2$ the kernel is too coarse to give as good results as 
for $N=1$. The learned classification boundaries are given as contour plots.
The slight difference in performance of KQML 
in relation to the theoretical prediction is due to statistical fluctuations of 
the experimental data and relatively small test set (misclassification of a single near-boundary point results in $0.02$ performance drop) .}
\vspace*{-3mm}
\end{figure}

%\subsection{Experimental implementation}
\paragraph*{Experimental implementation.}
We have experimentally implemented KQML to solve three classification 
problems on a two-photon optical quantum computer. In our experiment we implemented 
a $D=2,N=1$ kernel (using all the modes from Fig.~\ref{fig:2}, we can set at most $D=5$ with $q=1$). 
We used two photons, but only the top mode of the dual-rail encoding. 
Including more modes would lead to kernels causing overfitting (see Fig.~\ref{fig:3}).
We have performed measurements for $M=40$ two-dimensional samples ($D=2$), drawn
from two classes (see horizontally/vertically-tipped triangles in Fig.~\ref{fig:3}). 
This procedure was repeated for three benchmark classification problems. 
For each benchmark $40\times 39/2=780$ measurements 
were performed to create a corresponding 
Gramm matrix (GM), which was subsequently used to find the best linear classification 
boundary as given by the representer theorem. In other words, a custom kernel 
$\kappa(x^m,x^{n})=\kappa(x^n,x^{m})$ for $m,n=1,2,...,M$ was measured. 
This kernel was used as a custom precomputed kernel 
for the scikit-learn SVC classifier in python. 
Pairs of $H$-polarized photons were prepared in a type-I spontaneous 
parametric down-conversion process in a $\beta$--BaB${}_2$O${}_4$ crystal. The crystal was 
pumped by a $200$\,mW  laser beam at 355\,nm
(repetition rate of 120\,MHz). The coincidence rate, 
including all possible detection events from Fig.~\ref{fig:1}, was 
approximately $250$ counts per second.  The setup operates with high fidelity (98\%)
and the dominant source of errors is the Poissonian photon count statistics. 
We measured each point for a time necessary to collect about 
$2500$ detection events. Thus, 
excluding the time needed to switch the setup parameters, the whole measurement 
for a single benchmark problem takes about two hours. 
To prepare the contour plot of the decision function based on the experimental 
data shown in Fig.~\ref{fig:3} and to quantify the performance of the trained model
on the relevant test sets, we have also measured the GM for 
$1225$ points and used its symmetries to fill in the unmeasured values. 
The values for points in between have been found using linear interpolation. 
The data accumulation time can be shortened by orders of magnitudes 
by fine tuning the parameters of the setup and by using brighter photon sources.

%\section{Conclusions}
%\label{conclusions}
\paragraph*{Conclusions.}
We report on the first experimental implementation of 
supervised QML for solving a nonlinear multidimensional
classification problem with clusters of points which are not
trivially separated in the feature space.  
We hope that our research on QML will help to improve ML technologies, which 
are a major power-horse of many industries, a vivid field of research in 
computer science, and an important technique for solving real-world problems. 
We believe that both the theoretical and the experimental 
investigation of FM circuits and their constraints regarding kernel 
resolution and compression for a limited FHS (i.e., FHS size dependent 
FMs)  constitutes a crucial step in the development 
of practical KQML for SVM QML \cite{Rebentrost2014quantum,PhysRevLett.114.140504,Chatterjee2017,Schuld2019PRL,havlivcek2019supervised}. 
We demonstrate that a linear-optical setup with discrete photon encoding 
is a reliable instrument for this class of quantum machine learning tasks.
We also report obtaining exponentially better 
scaling of FHS in the case of CK
than in the case of taking direct products of qubits~\cite{Schuld2019PRL}. 
The same can hold for other more complex kernels implemented
in finite FHS, which could appear unfeasible, but in fact require more elaborate FMs
(e.g., the resolution-optimized kernels shown in Fig.~\ref{fig:1}).
Thus, KQML can provide a promising perspective for utilizing noisy intermediate-scale quantum systems~\cite{Preskill2018quantumcomputingin,PhysRevApplied.8.024030,PhysRevLett.121.250501,kandala2018extending}, complementing artificial quantum neural networks~\cite{shen2017deep,Bueno:18,tacchino2019artificial,kak1995quantum,farhi1802classification} and other hybrid quantum-classical algorithms~\cite{PhysRevX.7.021050,kandala2017hardware,PhysRevA.98.032309}.

%\section*{Acknowledgement}
\paragraph*{Acknowledgements}
\begin{acknowledgments}
Authors acknowledge
financial support by the Czech Science Foundation under the project No. 
19-19002S. The authors also acknowledge the projects Nos. LO1305 and 
CZ.02.1.01./0.0/0.0/16\textunderscore 019/0000754 of the Ministry of Education, 
Youth and Sports of the Czech Republic financing the infrastructure of their 
workplace. F.N. is supported in part by the: 
MURI Center for Dynamic Magneto-Optics via the 
Air Force Office of Scientific Research (AFOSR) (FA9550-14-1-0040), 
Army Research Office (ARO) (Grant No. Grant No. W911NF-18-1-0358), 
Asian Office of Aerospace Research and Development (AOARD) (Grant No. FA2386-18-1-4045), 
Japan Science and Technology Agency (JST) (via the Q-LEAP program, and the CREST Grant No. JPMJCR1676), 
Japan Society for the Promotion of Science (JSPS) (JSPS-RFBR Grant No. 17-52-50023, and JSPS-FWO Grant No. VS.059.18N), 
and the RIKEN-AIST Challenge Research Fund.
\end{acknowledgments}

\end{document}